\pdfoutput=1
\documentclass{AISB2008}
\usepackage{times}
\usepackage{graphicx}
\usepackage{latexsym}

\begin{document}

\title{Faith in the Algorithm, Part 1: \\ Beyond the Turing Test}

\author{Marko A. Rodriguez\institute{Theoretical Division -- Center for Non-Linear Studies, Los Alamos National Laboratory, email: marko@lanl.gov}  \and Alberto Pepe\institute{Center for Embedded Networked Sensing, University of California, Los Angeles, email: apepe@ucla.edu} }

\maketitle

\begin{abstract}
Since the Turing test was first proposed by Alan Turing in 1950, the primary goal of artificial intelligence has been predicated on the ability for computers to imitate human behavior. However, the majority of uses for the computer can be said to fall outside the domain of human abilities and it is exactly outside of this domain where computers have demonstrated their greatest contribution to intelligence. Another goal for artificial intelligence is one that is not predicated on human mimicry, but instead, on human amplification. This article surveys various systems that contribute to the advancement of human and social intelligence.
\end{abstract}

\vspace*{0.0in}
\begin{quote}
\textit{The alleged short-cut to knowledge, which is faith, is only a short-circuit destroying the mind.} \begin{flushright}-- Ayn Rand, ``For the New Intellectual''\end{flushright}
\end{quote}
\vspace*{0.0in}

\section{INTRODUCTION}

The path towards artificial intelligence, in terms of mimicking human cognitive functionality, has been long, difficult, and painfully incremental. Bottom-up, state of the art vision systems have only accomplished modeling the functional capabilities of the V1, V2, and V4 regions of the visual cortex \cite{network:serre2007}. Popular, top-down knowledge representation and reasoning system are still primarily monotonic \cite{owlspec:mcguinness2004}, are only beginning to incorporate and understand the ramifications of common sense knowledge \cite{muller:common2006}, and are predicated on logics that do not appear to model the true ``rules'' of human thought \cite{coglogic:wang2004}. Moreover, these object recognition and knowledge representation and reasoning developments are but the fringe of a huge landscape of cognitive faculties that must be simulated to achieve human-type artificial intelligence in its fullest form. For example, other less developed agendas are object relation learning in neurally-plausible substrates \cite{hummel:symbolconn2001}, novel logic acquisition through experience \cite{nal:wang2006}, and associative mechanisms for merging the categorizations from different sensory modalities into a single language of thought \cite{lot:fodor1975,intell:hawkins2005}. 

The sub-symbolic agenda of artificial intelligence attempts to model the lowest common denominator of the human neural system in order to achieve higher levels of intelligence through experience and learning. Modeling the processing capabilities of individual neurons has been the aim of the connectionist agenda for nearly three decades \cite{rumelhart:conn1993} and beyond various advances in classification, it appears that human type intelligence is still many more decades away. In the area of symbolic artificial intelligence, there have been many developments utilizing computers to solve very specific problems very well, but unfortunately, many of these systems do not have the general, flexible intelligence enjoyed by humans. These statements serve not to criticize the researchers or their methods; rather, they are presented in order to acknowledge the level of difficulty involved in simulating human-type intelligence and the distances that need to be reached if this goal is to be achieved. Is it possible that computers, and their underlying foundation in bivalent logic, centralized processing, and disembodiment, are blinding us as architects and engineers by biasing our approach \cite{clark:beingthere1997}? Of course, this does not mean that it is impossible to model human intelligence on a computer (assuming that such intelligence can be modeled on a Turing complete system). Instead, it is more a statement that the Turing test \cite{turing:test1950} -- the test for computer intelligence by means of human mimicry -- is not a ``natural'' test of the computer's abilities in the area of intelligence. Moreover, human mimicry is not a ``natural" application of the computer's abilities.

There are many tests that are used to quantify human intelligence. Interestingly, in the mean, a human subject's scores in all of these tests have a positive correlation. Thus, regardless if a specialist is testing a subject's ability to manipulate objects in 3D space or the subject's fluency with language, success in one of these tests is a predictor of success in another. This finding points to a single factor that can account for intelligence. This factor is known as the $g$-factor (or general intelligence factor) \cite{spearman:g1904}. However, any test for intelligence ultimately makes assumptions about the sense modalities through which the test will be administered as well as assumptions about the cultural and common knowledge of the subject. A major trend in intelligence test research is to make intelligence tests devoid of any cultural biases and one day, it may be possible to yield tests that are devoid of any species and modality biases. Species agnostic intelligence tests could be used to measure the intelligence exposed at the level of the human/computer as the autonomous, intelligent entity. Moreover, the degree of intelligence may be greater than what is possible given the human or computer alone \cite{supersize:clark2008}. This is because the computer demonstrates unmatched skills in very specific areas such as quickly computing the distance between large vectors of numbers or in maintaining a lossless representation of a presented image in memory. Such skills and their relationship and integration with the skills of the human will continue to yield an advanced degree of real-world intelligence. It is the central thesis of this article that this contribution to intelligence appears to be a more ``natural'' fit for the computer. This article reviews various systems that, when in combination with humans, yield advanced intelligence -- an intelligence that is different than that which can be exposed by the Turing test.

\section{HUMAN AND SOCIAL AUGMENTATION}

Computers -- the machines and their implemented algorithms -- should not simply be interpreted as technological embodiments of solutions to specific problems. There is a larger relationship between the human, their problems and requirements, and designed algorithms and their executing hardware. They are solving larger problems than either the human or the computer could solve alone; in other words, the computer is a contributing component within a larger intelligent system \cite{heylighen:gb2007}. Sherry Turkle discusses the relationship between humans and computers as not just one in which the computer is a tool used to accomplish human tasks, but one where it is  a component that works within the human's everyday life as a supporting entity \cite{secondself:turkle1984}. From a ``society of minds'' perspective \cite{minsky:som1988}, the computer, as a cognitive component in human thinking, is very much a well functioning digital information processor much like the hippocampus is a well functioning neural memory device. In other words, the computer has found, not in any affective directed way, an information processing niche that further augments the human much like any other component of the human neural system \cite{skagestad:augment1993}. To say whether the hippocampus is intelligent or not is to determine whether the results of its processing affect intelligent behavior; that is, does the human know where they are in physical space and do they encode episodic memories correctly? As an autonomous entity, the hippocampus, would appear, to the external human observer, as not being intelligent at all. For one, in isolation, it simply becomes infected and its cells quickly die. However, within the larger schema of the human organism, its role is of great significance to human intelligence. A few minutes interaction with the patient H.M. makes this point obvious \cite{hm:cohen1995}. Next, looking at the striate cortex demonstrates a relatively simple system \cite{hubel:striate1968} that implements a relatively simple algorithm (albeit on a massive scale) \cite{network:serre2007}; however, when integrated within the nervous system as a whole, the contribution of the striate cortex to the overall intelligence of the human is immense. Without it, vision, and its associated functionalities, would not be possible. For instance, there would be no notion of a genius painter and the level of intelligence that such a connotation denotes. To this end, how many neural components are required before it is assumed that a human is intelligent? A review of the life and times of Helen Keller should demonstrate how vacuous this question is \cite{helen:keller1905}. Also, like the neural component within the larger system of the human, any other processing component can be utilized in this contribution to intelligence. As such, the measurement of intelligence need not be considered as testing that which is within the confines of the human skin.


The relationship between the human and the computer in a technologically-driven society unveils a natural symbiosis which is reminiscent of Hutchins' theory of distributed cognition \cite{hutchins:congition1995} and to the notions of collective intelligence found in ant and termite populations \cite{grasse:stig1959,bonabeau:swarm1999}. Some of the tasks in which computers are employed in everyday life -- from information access to social interaction -- make this symbiosis evident. In many respects, traditional, standardized tests of human intelligence test the emergent behavior of the coordinated activity of the individual's various brain regions. Introducing the computer into this system simply augments or extends the intelligent capabilities of the individual human. It is no accident that this symbiosis has emerged. The computer and its associated algorithms is a needed augmentation to the human given the number of options available in the technologically-rich world and the difficulties in finding one's global optima within it. Moreover, society, in a collaborative fashion amongst its constituents and its supporting digital infrastructure, is making and will continue to make advances in the area of social intelligence. In this light, the question at hand is: what is the computer's contribution to intelligence? In order to address this question, the following section explores the emergence of advanced individual and social intelligence within the scope of the technological innovation that has most contributed to this type of augmentation in recent times: the World Wide Web.

\section{EMERGENT WEB INTELLIGENCE}

Since the dawn of the World Wide Web, information has been codified and distributed within a shared, universal medium that is accessible by human users world wide. The World Wide Web is unique for two reasons: distribution and standardization. In many respects, the first can not be accomplished without the latter. The Web's most eminent standard, the Uniform Resource Identifier (URI) has made it possible for the Web to serve as a network of information, from the document to the datum -- a shared, global data structure \cite{pubsem:lee2001}. This distributed data structure is amplifying the intelligence of the individual human and may provide a greater social intelligence. The remainder of this section will address the amplification of intelligence in the context of three general Web system: search engines (index and ranking), recommendation engines (personalized recommendations), and governance engines (collective decision making) \cite{webci:watkins2007}.

\subsection{Search Engines}

The World Wide Web has emerged as a massive information repository from which humans contribute and consume information. This has not only provided a simple means of retrieving information, but also a simple way to publish and distribute information, thus leading to the increase in human information production. However, information increase inevitably brings about discoverability issues, as the necessity to locate and filter desired information arises. To deal with this problem, algorithms have been developed to augment the individual's search capabilities. Interestingly, this augmentation is currently predicated on the contribution of many individuals within the stigmergetic environment of the World Wide Web.

The early Web maintained rudimentary indexes in the form of Web ``yellow pages'' that provided short descriptions of web pages. With the explosive growth of the Web, such directory services fell by the wayside as no human operator (or operators) could keep up with the amount of information being published, nor could such rudimentary lists provide the end user a representation of the quality of web pages. By a nearly-Darwinian selection process, these early forms of indexes fell out of use because they were built around a conceptual framework that did not take advantage of the distributed representation of value inherent in every linking webpage made explicit by their authors. As a remedy to this situation, a commercialized Web industry was born and continues to thrive around solving the problem of search. Search engines index massive amounts of data that are gleaned from Web servers world wide. The development of the simple mechanism of ranking web pages by means of their eigenvector component within the web citation graph has proved the most successful to date \cite{anatom:brin1998}. It is remarkable that this mechanism is predicated on humans' decisions to link webpages; that is, the algorithm leverages human interaction with the Web and vice versa in a symbiotic manner. Even more remarkable is the fact that with the approximately 30 billion web pages in existence today, Web users can rest assured that, for the most part, their keyword search will provide the answer to their question within the first few results returned. This level of speed and accuracy of knowledge acquisition was not possible prior to the development of the Web, mainly because the problem of massive-scale indexing and ranking did not make itself apparent until the Web. This problem is solved through the unification of the human's ability to, in a decentralized fashion, denote the value (or quality) of web pages and the computer's ability to calculate a global rank over these explicit expressions of value.
 
In this scenario, the Web plays the role of a digital Rolodex providing the human, nearly instantly, a reference to further information on nearly any topic imaginable \cite{engelbart:memex1988}. Prior to the written document, information was passed from generation to generation in the form of large memorized stories and poems. In the contemporary technologically-rich world,  this ``algorithm'' (cultural process) is no longer necessary. This is not to say that an individual can no longer memorize a long poem if they wish. It is more that a new algorithm has emerged to handle this information indexing requirement and as such, cognitive resources can be appropriated to other tasks. However, the Web is not a large story or poem: it follows no plot, no linear sequence, no poetic meter, no single language -- the list of characters is beyond count and no one writing style can be identified. For these reasons, it is posited that no currently existing neural component can memorize, index, and rank the entire Web, and thus, a specialized intelligence is required and, fortunately, has emerged.

\subsection{Recommendation Engines}

Large-scale human generated data sets have opened a terrain for numerous algorithms that support individual decision making. Such data sets include the implicit valuation of resources that users leave on the web as they click from web page to web page or from purchased item to purchased item. No individual ever sees the entire Web and for the most part, for the life of the individual, they are confined to a small subset of the greater Web. However, the aggregation of this click-stream information from all individuals provides a collectively generated representation of the inherent relationship between all items on the Web. This collective digital footprint provides not only novel ways to rank resources \cite{bollen:mesur2008} but also, novel ways to recommend resources \cite{video:bollen2007}. Finally, humans are also developing rich profiles of themselves that include not only identifiable facts such as one's curriculum vitae, but also the more qualitative aspects of their personality, tastes, and ever changing mood. There are many systems that take advantage of such data sets such as the recommendation engine. A recommendation engine can be defined as any algorithm that provides users with resources (e.g.~documents, books, music, movies, life partners, etc.) that are more likely than not to be correlated to the users' current requirements. 

The popular collaborative filtering algorithms of document and music services are able to utilize the previous click behavior of an individual to systematically compare it with the click behaviors of others, and from this comparison, recommend a set of resources that will be of interest to the user \cite{collab:herlocker2006}. For many, the dependency on the librarian and the record shop owner has shifted to a dependence on the community as a whole that is leaving this massive digital footprint.

An interesting phenomena to arise in recent years is the development and use of online dating services. In any large city, there are too many individuals for any one human to sift through. Moreover, even if an individual were able to meet everyone, the abilities of the individual may not be keen enough to predict, with any great accuracy, whether or not the person they are meeting will make an optimal partner. For this reason, dating services have emerged to handle, or rather attempt to handle, this common, pervasive problem. Ignoring broader social and cultural considerations for a moment, from a purely statistical perspective, the human's trial and error methods of sampling small portions of the population through friends or in social, physical environments (bars, restaurants, cafes, etc.) can not compete with the success rates of modern day matchmaking algorithms \cite{love:aaron}. Note that matchmaking services are not confined solely to the Web. Newspapers provide ``personals'' sections, but like the early ``yellow pages'' of the Web, they can not maintain rich profiles, nor does manually browsing this information compare with the success of a matchmaking algorithm's recommendation. Again, for those activities for which a human simply does not have the skills to succeed, the human relies on an external augmentation to fulfill the intelligence requirements of the problem at hand.

Recommendation services are following a common trend: they are all building more sophisticated models of both humans and resources. The World Wide Web infrastructure has provided the avenues for humans to collectively aggregate in a shared virtual space. Unfortunately, for the most part, the traffic data that is being generated as individuals move from site to site, the profiles that individuals repeatedly create at every online service, and the metadata about the resources that these services index are isolated within the data repositories of the services that utilize this information directly. Fortunately, recent developments in an open data model known as the ``web of data'' may change this by unifying the information contained in service repositories and exposing, within the shared, global URI address space, every minutia of data \cite{linkeddata:bizer2008}. The end benefit of this shift in the perception of ownership and exposure of data will allow for a new generation of algorithms that take advantage of an even richer world model \cite{lessig:creative2008,rodriguez:distributed2008}. Such models will include a seamless integration of the individual's reading, listening, dating, working, etc. behaviors as well as the descriptions of books, songs, movies, people, jobs, etc. At this point, to the algorithms that leverage such data, a human is no longer just a consumer of a particular type of literature or a connoisseur of a particular style of film, but rather, a complex entity that can be subtly oriented, through recommendation, in a direction that ensures that they are experiencing that aspect of the world that is most fitting to who they are.

At the extreme of this line of thought, if enough information is gathered and a rich enough world model is generated, then it may be possible to design algorithms that are more fit to determine the life course of an individual human than what the individual, their family, or their community can do for them. This assumes appropriate feedback from the world to the model \cite{vadas:inter2007}, which may include the perspectives of the individual, their family, and their community. This view suggests that it may be best to rely on a large-scale world model (and algorithms that can efficiently process it) when making decisions about one's path in life. Such algorithms can take into account the multitude of relations between humans and resources, and improvise a well ``thought out'' plan of action that ensures that the individual, to the best of the system's ability, lives a life that is filled with optimal experiences. This is a life in which the others they meet, the restaurants they frequent, the books they read, the classes they attend, and so forth lead to experiences that are completely fulfilling to them as a human. These optimal experiences represent the perfect balance between the psychological states of anxiety and boredom and as such, would increase the individuals' attentiveness and involvement in such activities -- similar to the mental state that is colloquially known as ``flow'' \cite{csi:flow1990}. Moreover, this state of human experience has been articulated since the times of Aristotle and his notion of the eudaemonic living which arises when one consistently chooses correctly in their life \cite{nico:aristotle1998}. 

A large-scale world model has the potential to integrate the collective zeitgeist of a society, the socio-demographic and geographic layouts of cities, the location of its inhabitants, their personal characteristics, their resources and relations. Amazingly, such data currently exist in one form or another, to varying degrees of accuracy, completeness, and levels of access. Further making this information publicly available and integrated would allow for algorithms to evolve, over iterations of development and insight, that are fit to determine the individuals' global optima.

\subsection{Governance Engines}

In many ways, aiding the human in finding global optima is the purpose of a society (within the constraints of taking into account the optima of others) \cite{demo:norton1995}. From high-level governmental decisions to the local cultural rules that determine the way in which humans interact in their environment, the goal of a (benevolent) society is to ensure a life in ``the pursuit of happiness'' \cite{jefferson:declaration1776}. However, can a society be structured such that the individuals need not pursue, but instead be guaranteed a life full of happiness -- or eudaemonia and optimal experiences? The question is then: what are the limits of individual intelligence that can be achieved by the current societal structures alone? And also: are there more efficient and accurate algorithms that can be utilized? Recommendation systems are a step in the direction towards the use of computers to provide the human the right resource at the right time, regardless of what form that resource may take. However, within the grander scheme of society as a whole, the nascent fields of e-governance and computational social choice theory are only beginning to tangentially touch upon the idea that a networked computer infrastructure could be used to foster a new structure for government that is optimized for societal-scale problem-solving.

Reflecting on modern voting mechanisms (specifically those within the United States), we find a system that is fragile, inaccurate, and expensive to maintain. Due in part to the outdated infrastructure that citizens use to communicate with their governing body, citizen participation in government decision making is limited. However, these days, with the level of eduction that citizens have, the amount of information that citizens can become aware of, and the sophistication of modern network technologies, it is possible that current government decisions are limited in that they are not leveraging the full potential of an enlightened population (or subset thereof). By making use of both a large-scale and knowledgeable decision making constituency, it is theoretically possible that all rendered decisions are optimal. This statement was validated (under certain simple assumptions) in 1785 by Marquis de Condorcet's now famous Condorcet jury theorem \cite{condorcet:theorem1776}. 

With the social networks that are being made explicit on the Web today, and with open standard movements that ensure that this information can be shared across services, it is possible to leverage a relatively simple vote distribution mechanism to remove the representative layers of government and promote full citizen participation in all the decision making affairs of a society. This mechanism, known as dynamically distributed democracy, ensures that any actively participating subset of a population simulates the decision making behavior of the whole \cite{ddd:rodriguez2004}. Thus, a simulated, large-scale decision making body can be leveraged in all decisions. A large decision making body is the first requirement of the Condorcet jury theorem. Robin Hanson articulates a vision of government where any individual can participate through a decision system known as a prediction market \cite{futurarchy:hanson}. The purpose of a prediction market is to provide accurate predictions of objectively determinable states of the world (current or into the future) and its application to governance is noted in the popular phrase ``vote on values, but bet on beliefs." In this form, the self-selecting, monetary mechanisms that determine whether someone participates is based on their degree of knowledge of the problem space. Those that are not knowledgeable, either do not participate or lose money in the process of participating, thus, hampering the individual from participating in matters outside the scope of their abilities into the future. The accuracy of such systems are astounding and have popular uses in election predictions and a short lived run in terrorist predictions (only to be dismantled by the U.S. government because it was considered too morose for market traders to monetarily benefit on the accurate prediction of the death of others). A knowledgeable decision making body is the second requirement of the Condorcet jury theorem and, much like commodity markets, prediction market systems select for knowledgeable individuals.

These ideas stress the importance of reflecting on the medium by which society organizes itself, generates its laws, and implements methods in how it will utilize resources most effectively. Like the ``yellow pages'' of the early Web, it may not be optimal to leave such pressing matters to an operator (or operators). This statement is not a critique of the leaders and doctrines of nations, but instead is a comment on the complexity of the world and the necessity for a new type of intelligence. It is posed as an appeal to rethink government and its role within contemporary networked society \cite{enlighten:rodriguez2009}. An implementation of a government should not be valued. Instead, what should be valued is the ideals that that implementation is trying to achieve. Moreover, if another implementation would better meet the ideals of the society, then it should be enacted. A distributed value/belief system and algorithmic aggregation mechanism may prove to be the better problem-solving mechanism for societal issues and may prove to be a better mechanism to orchestrate individual lives. It is in this area that computers can greatly contribute to social intelligence, where the unification of the intelligence augmentation gained by the individual human and the society coalesce into a type of intelligence that is novel (beyond human mimicry) and above all beneficial.

\section{CONCLUSION}

Humans perceive their world through their sense modalities, create stable representations of the consistent patterns in the world, and utilize those representations to further act and survive to the best of their abilities. Their internal, subjective world is an endless stream of thoughts -- a complex, information-rich map of the external world. Manifestations of intelligence inherently depend upon an individual's internal representation of the external world and their ability to manipulate that representation. By analogy to the field of computer science, this internal map of the world can be regarded as the data structure upon which reasoning mechanisms (i.e.~algorithms) function. From an objective perspective, the human mind can only maintain so rich a data structure, process only so many aspects of it, and simulate only so many potential future paths for the individual to choose from. The complexity of the human's mental calculation grows when considering that many other such simulations are occurring in the minds of their fellow men and women. Like a general-purpose processor, to simulate a machine within a machine reduces the resources available to the original machine to execute other processes. For these reasons, the human is not a perfectly intelligent creature always doing the right thing at the right time.

As discussed, with the externalization of the human's internal world through the explicit expression of themselves, their relation to others, and the resources on which they rely, other processes can utilize this explicit model to aid the human in the process of thought and thus, life. The World Wide Web and the algorithms implemented upon it function like an auxiliary mind, exposed to more information than could be possibly processed by its neural counterpart. While the core specification of these algorithms may be understood, even thoroughly by their designers, ultimately what machines compute are based on such a large-scale model of the world, that to assimilate its results into one's choices are ultimately based on faith -- much like the faith one has in the validity of their episodic memories and their current location in space as provided to them by their hippocampus.

\end{document}